\documentclass[a4paper,fleqn,usenatbib]{mnras}
\usepackage[T1]{fontenc}
\usepackage{ae,aecompl}

\usepackage{amsmath}
\usepackage{graphicx}
\usepackage{lscape}
\usepackage{indentfirst}

\usepackage{amssymb}
\usepackage{color}
\usepackage[flushleft]{threeparttable}

\newcommand {\bc}{\begin {center}}
\newcommand {\ec}{\end {center}}
\newcommand {\be}{\begin {equation}}
\newcommand {\ee}{\end {equation}}
\newcommand {\beq}{\begin {eqnarray}}
\newcommand {\eeq}{\end {eqnarray}}

\newcommand {\ergs}{{\rm erg\ \rm s^{-1}}}
\newcommand {\comment}[1]{}

\def\intl {\int\limits}
\def\lbar {\lambda\hskip-5pt\raise3pt\hbox {--}}
\def\lbr {\lambda\raise2pt\hbox {\hskip-4pt{\scriptsize --}}_\C}

\renewcommand{\d}{{\rm d}}

\renewcommand{\d}{{\rm d}}

\title[Timing properties of ULX pulsars]
{Timing properties of ULX pulsars: optically thick envelopes and outflows}
\author[A. A.~Mushtukov et al.] 
{Alexander~A.~Mushtukov,$^{1,2,3}$\thanks{E-mail: al.mushtukov@gmail.com (AAM)}  
Adam Ingram,$^{4}$
Matthew Middleton,$^{5}$ \newauthor
Dmitrij~I.~Nagirner,$^{6}$ and Michiel van der Klis$^{1}$ \\ 
$^1$ Anton Pannekoek Institute, University of Amsterdam, Science Park 904, 1098 XH Amsterdam, The Netherlands \\
$^2$ Leiden Observatory, Leiden University, NL-2300RA Leiden, The Netherlands \\
$^3$ Space Research Institute of the Russian Academy of Sciences, Profsoyuznaya Str. 84/32, Moscow 117997, Russia \\
$^4$ Department of Physics, Astrophysics, University of Oxford, Denys Wilkinson Building, Keble Road, Oxford OX1 3RH, UK \\
$^5$ Department of Physics and Astronomy, University of Southampton, Highfield, Southampton SO17 1BJ, UK \\
$^6$ Sobolev Astronomical Institute, Saint Petersburg State University, Saint-Petersburg 198504, Russia \\
}

\pubyear{2018}

\begin{document}
\label{firstpage}
\pagerange{\pageref{firstpage}--\pageref{lastpage}}
\maketitle

\begin{abstract}
It has recently been discovered that a fraction of ultra-luminous X-ray sources (ULXs) exhibit X-ray pulsations, and are therefore powered by super-Eddington accretion onto magnetized neutron stars (NSs). For typical ULX mass accretion rates ($\gtrsim 10^{19}\,{\rm g\,s^{-1}}$), the inner parts of the accretion disc are expected to be in the supercritical regime, meaning that some material is lost in a wind launched from the disc surface, while the rest forms an optically thick envelope around the NS as it follows magnetic field lines from the inner disc radius to the magnetic poles of the star. The envelope hides the central object from a distant observer and defines key observational properties of ULX pulsars: their energy spectrum, polarization and timing features. The optical thickness of the envelope is affected by the mass losses from the disc. We calculate the mass loss rate due to the wind in ULX pulsars, accounting for the NS magnetic field strength and advection processes in the disc. We argue that detection of strong outflows from ULX pulsars can be considered evidence of a relatively weak dipole component of the NS magnetic field. We estimate the influence of mass losses on the optical thickness of the envelope and analyze how the envelope affects broadband aperiodic variability in ULXs. We show that brightness fluctuations at high Fourier frequencies can be strongly suppressed by multiple scatterings in the envelope and that the strength of suppression is determined by the mass accretion rate and geometrical size of the magnetosphere.
\end{abstract}



\section{Introduction}

Ultraluminous X-ray sources (ULXs) are unresolved extra-galactic off-center sources of X-ray luminosity $L>10^{39}\,\ergs$ \citep{2017ARA&A..55..303K}, which is already above the Eddington luminosity for accreting neutron stars (NSs): 
$L_{\rm Edd}\approx 1.8\times 10^{38}(M/1.4M_\odot)\,\ergs$. 
For a long time, most theories explaining ULXs were focused on super-critical accretion onto stellar mass black holes (BHs, \citealt{2006MNRAS.370..399B,2007MNRAS.377.1187P}) 
or sub-critical accretion onto intermediate mass BHs \citep{1999ApJ...519...89C,2017mbhe.confE..51K}. 
However, it has recently been discovered that some ULXs show coherent pulsations  \citep{2014Natur.514..202B,2016ApJ...831L..14F,2017MNRAS.466L..48I,2017Sci...355..817I} and are, therefore, extreme cases of X-ray pulsars (XRPs, see e.g. \citealt{2015A&ARv..23....2W}) - ULX pulsars - powered by super-Eddington accretion onto magnetized NSs (as predicted by \citep{2001ApJ...552L.109K}).
There are 4 known ULX pulsars discovered to-date:
ULX M82 X-2 \citep{2014Natur.514..202B},
NGC 7793 P13 \citep{2016ApJ...831L..14F,2017MNRAS.466L..48I},
ULX-1 in the galaxy  NGC 5907 \citep{2017Sci...355..817I,2017ApJ...834...77F},
and ULX-1 in the galaxy NGC 300 \citep{2018MNRAS.476L..45C}. 
Recently, a probable detection of a narrow cyclotron scattering feature at $\sim 4.5\,{\rm keV}$ has been reported for ULX-8 in the galaxy M51 \citep{2018NatAs...2..312B}. Pulsations were not detected in this particular ULX, but the feature was interpreted as a proton cyclotron line, implying a surface multipole field of $\sim 10^{15}\,{\rm G}$ and making this source the fifth candidate ULX pulsar.
The detected X-ray luminosity of the brightest ULX pulsar discovered to date reaches $\sim 10^{41}\,\ergs$ \citep{2017Sci...355..817I} and, thus, exceeds the Eddington luminosity for a NS by a factor of a few hundred.

The nature of the enormous luminosity of ULX pulsars is still under debate.
Different models consider a wide range in NS magnetic field strength:
from $\sim 10^{11}\,{\rm G}$ at the stellar surface, with strong outflows and geometrical beaming (see e.g. \citealt{2017MNRAS.468L..59K}), up to magnetar-like fields of $\sim 10^{14}\,{\rm G}$ \citep{BS1976,2015MNRAS.448L..40E,2015MNRAS.449.2144D,2016MNRAS.457.1101T}, which are strong enough to significantly reduce radiation pressure and confine accretion columns above the NS magnetic poles \citep{2015MNRAS.454.2539M}. A number of additional factors may be essential to the physics of ULX pulsars: a complicated $B$-field geometry \citep{2017Sci...355..817I,2018MNRAS.479L.134T}, variability in geometry of the accretion flow \citep{2017AstL...43..464G}, photon bubbles \citep{1992ApJ...388..561A,2006ApJ...643.1065B}, and strong neutrino emission from advection dominated accretion columns \citep{2018MNRAS.476.2867M}.

It has been shown that ULX pulsars should be surrounded by optically thick (for Compton scattering of X-ray photons by electrons) envelopes formed by material at the NS magnetosphere free-falling from the accretion disc to the central object \citep{2017MNRAS.467.1202M}. The envelope is expected to be optically thick for a wide range of values for the magnetic field strength ($\lesssim 10^{14}\,{\rm G}$) and for any reasonable beaming factor ($b\le 10$), and may be a key ingredient providing the principal possibility] in the mechanism of matter transfer from the accretion disc to the NS.
The envelope hides the central object from a distant observer and shapes key observational properties of ULX pulsars: smooth pulse profiles, relatively soft X-ray energy spectra and undetectable (in all pulsating ULXs discovered up to day) cyclotron lines, which are expected to be very weak due to X-ray spectra being strongly modified by Comptonization in the envelope. 
The appearance of optically thick envelopes was shown to be in a good agreement with observations \citep{2017A&A...608A..47K}.

Appearance of the envelope and reprocessing of X-ray photons in it should also influence the typical time scale of photon escape from the source
\citep{1983ApJ...265.1005C}: X-ray photons originating from the central engine \citep{BS1976,2015MNRAS.454.2539M,2018MNRAS.474.5425M} experience a number of scatterings before leaving the envelope. 
As a result, any variability of X-rays on time scales smaller than the typical time scale of photon escape is expected to be smoothed out.
Accreting NSs are known to be sources of a strong aperiodic variability over a very large frequency range extending up to hundreds Hz \citep{1993ApJ...411L..79H,2006csxs.book...39V}.
Aperiodic variability of the X-ray energy flux from XRPs arises from variability of the mass accretion rate in the very vicinity of the NS surface \citep{2009A&A...507.1211R}, which is a result of mass accretion rate fluctuations arising throughout the accretion disc and propagating inwards due to the process of viscous diffusion (see e.g. \citealt{1997MNRAS.292..679L,2018MNRAS.474.2259M,2018prep}).
The aperiodic variability at high Fourier frequencies in ULX pulsars should be strongly influenced by the envelope 
and, thus, analysis of rapid variability properties may enable verification of the whole concept of ULX pulsars hidden from the observer by the envelope. 

The geometrical size and optical thickness of the envelope are determined by the NS magnetic field strength and the mass accretion rate at the magnetosphere of the ULX pulsar. The mass accretion rate at the magnetosphere can be reduced with respect to the mass accretion rate from the donor star by a strong outflow from the disc \citep{1973A&A....24..337S}. 
In this paper, we construct a simple model of ULXs powered by accretion onto strongly magnetized NSs. 
Accounting for the possibility of strong outflows from super-Eddington advective accretion discs we re-estimate the geometrical size of the envelope and its optical thickness as a function of the mass accretion rate from the donor star and NS magnetic field strength.
Considering a toy model with a spherical envelope of a given optical thickness and a given geometrical size, and 
solving numerically the equations of radiative transfer in the envelope, we obtain constraints on the timing properties of X-ray radiation escaping from the system.

\section{The structure of the accretion flow in ULX pulsars}
\label{sec:AccFlowStructure}

\begin{figure*}
\centering 
\includegraphics[width=13.cm]{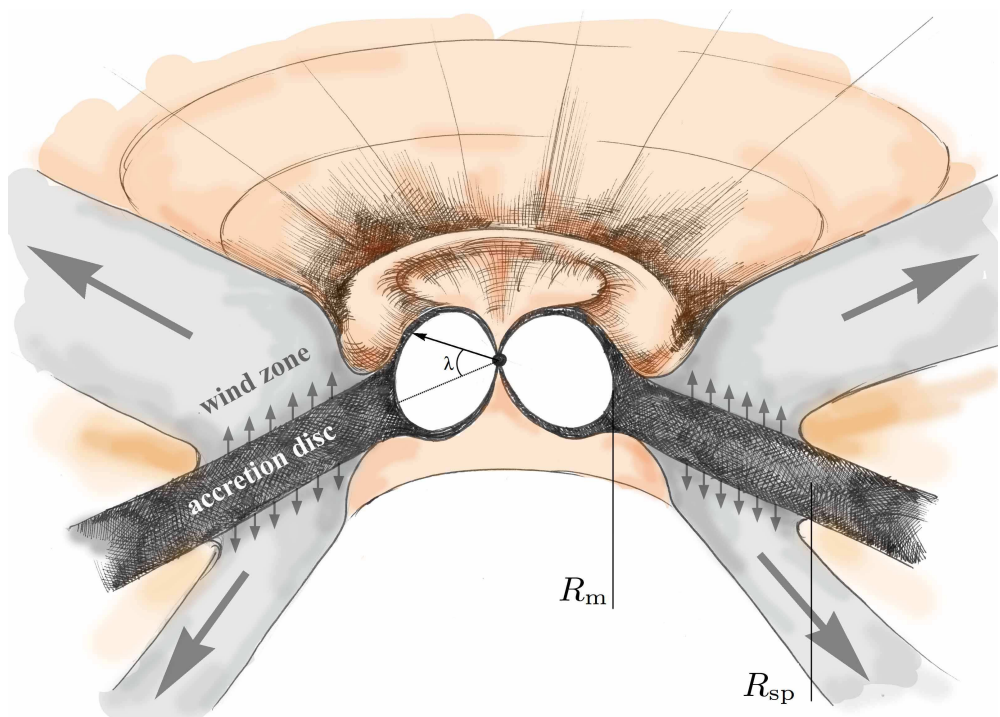}
\caption{
Schematic picture of a ULX powered by accretion onto a strongly magnetized NS.
The accretion disc is truncated from the inside at the magnetospheric radius $R_{\rm m}$ due to interaction with the magnetosphere of the NS.
If the mass accretion rate is high enough, the spherisation radius $R_{\rm sp}$ exceeds the magnetospheric radius and the accretion disc looses a fraction of matter due to the wind.
The rest of the matter accretes onto the central object forming an envelope, which tends to be optically thick at accretion luminosities typical for ULX pulsars.
}
\label{pic:ULX}
\end{figure*}

The accretion flow in X-ray binaries hosting highly magnetized NSs is truncated from inside due to interaction with the magnetic field of the central object (see Fig.\,\ref{pic:ULX}). 
If the accretion flow is truncated within the co-rotation radius
\be 
R_{\rm c}\approx 1.7\times 10^8 M_{1.4}^{1/3}P_{\rm s}^{2/3}\,\,{\rm cm},
\ee
where $M = M_{1.4}$ is the NS mass in units of $1.4M_\odot$ and $P_{\rm s}$ is the NS spin-period in seconds,
the accretion flow penetrates through the centrifugal barrier. In this case, the accreting plasma settles onto magnetic field lines \citep{2014EPJWC..6401001L} and moves towards the NS magnetic poles to form an envelope that becomes optically thick if the mass accretion rate is extremely high \citep{2017MNRAS.467.1202M}.
The size of the envelope and its optical thickness are determined by the mass accretion rate and the strength of the dipole component of the NS magnetic field.
The simplest estimation of the geometrical size is given by the magnetospheric radius
\beq 
\label{eq:Rm}
R_{\rm m}\approx 5.6\times 10^{7}\,\Lambda {B}_{12}^{4/7}\dot{M}_{19}^{-2/7}
M_{1.4}^{-1/4}R_6^{12/7}\,\,\,{\rm cm},
\eeq 
where $B_{12}$ is the magnetic field strength at the NS surface in units of $10^{12}\,{\rm G}$,
$\dot{M}_{19}$ is the mass accretion rate at the inner radius of the disc in units of $10^{19}\,{\rm g\,s^{-1}}$, and $R_6$ is the NS radius in units of $10^6\,{\rm cm}$.
The constant $\Lambda$ depends on the geometry of the accretion flow and its physical conditions: temperature, viscosity and ionization state. A value of $\Lambda=0.5$ is traditionally used for the case of a hot, geometrically thin accretion disc, whereas $\Lambda=1$ is appropriate for spherically symmetric accretion from the stellar wind \citep{GL1978,GL1992}. For disc accretion with extremely high mass accretion rates, the inner parts of the disc become radiation pressure dominated \citep{1973A&A....24..337S} meaning that the inner disc radius can be affected by the internal radiation pressure gradient and by radiative stress from the central object, resulting in an increase of $\Lambda$ up to unity (see Table 1 in \citealt{1999ApJ...521..332P}, see also \citealt{GL1992,2017MNRAS.470.2799C} for details). 

An even larger correction to the inner disc radius $R_{\rm m}$ at high mass accretion rates may arise from the fact that the disc losses a fraction of accreting material through an outflow driven by locally super-Eddington flux at the accretion disc surface. 
The outflow reduces the mass accretion rate in the inner parts of the disc and, thus, affects both the inner disc radius (see equation \ref{eq:Rm}) and the optical thickness of the envelope, which is formed by mass inflow from the inner disc radius to the central object. 

The envelope plays a key role in the accretion process at extreme mass accretion rates \citep{2017MNRAS.467.1202M}. 
Multiple scatterings of X-ray photons within the envelope result in thermalisation of the internal radiation.
As a result, the radiative stress opposes the magnetic pressure rather than the gravitational attraction of the central object, enabling mass transfer from the disc to the NS surface. 

\section{The influence of the outflow on the size and optical thickness of the envelope}
\label{sec:MassLosses}

At super-Eddington mass accretion rates, radiation pressure gradient in the inner parts of the disc become high enough to compensate gravitational attraction in the direction perpendicular to the disc plane.
The accretion disc becomes geometrically thick and able to produce winds driven by radiation force, spending a fraction $\varepsilon_{\rm w}\in[0;1]$ of viscously dissipated energy to launch the outflows.
As a result, only a fraction of the mass accretion rate from the donor star reaches the inner disc radius and accretes onto the central object.

Radiation pressure and outflows become important within the spherization radius, inside of which the radiation force due to the energy release in the disc is no longer balanced by gravity. This can be roughly estimated as \citep{1973A&A....24..337S,1999AstL...25..508L,2007MNRAS.377.1187P}:
\beq
R_{\rm sp}\approx 9\times 10^5\,\dot{m}_0\, 
\left[ 1.34-0.4\varepsilon_{\rm w}+0.1\varepsilon_{\rm w}^2 \right. \nonumber \\ 
\left. - (1.1-0.7\varepsilon_{\rm w})\dot{m}_0^{-2/3} \right]\,\, {\rm cm},
\eeq
where $\dot{m}_0=\dot{M}_0/\dot{M}_{\rm Edd}$ is the dimensionless mass accretion rate from the donor star in units of Eddington mass accretion rate at the NS surface calculated under the assumption of opacity dominated by Thomson scattering:
\be 
\dot{M}_{\rm Edd}=\frac{R\,L_{\rm Edd}}{GM}\approx 1.9\times 10^{18}\,R_{6}\,\,\,{\rm g\,s^{-1}}.
\ee
In the case of accreting BHs with an accretion disc extending down to the innermost stable orbit (ISCO) and NS with low magnetic field
\footnote{See numerical simulations performed by \citealt{2005ApJ...628..368O,2007PASJ...59.1033O,2017ApJ...845L...9T} for the case of accretion onto BHs and NSs with low magnetic field.}
, the outflow can carry out a significant amount of material from the accretion flow. 
In the case of accretion onto highly magnetized NSs, the inner disc radius can be much larger than the radius of the ISCO (see equation \ref{eq:Rm}) and the fraction of material carried out by the outflow is dependent both on the mass accretion rate and the magnetic field strength of the central object. 

\subsection{The inner disc radius and mass accretion rate onto the central object}

Let us estimate the outflow rate in the case of accretion onto a magnetized NS accounting for the $B$-field strength of the NS and the radial dependence of mass accretion rate. 
We follow the approximations obtained by \citealt{1999AstL...25..508L} and \citealt{2007MNRAS.377.1187P} for the case of accretion onto BHs and modify these calculations accounting for the truncation of the accretion disc at the magnetospheric radius (equation \ref{eq:Rm}), which depends on the mass accretion rate at the inner disc radius and the strength of the dipole component of the stellar magnetic field.

For super-Eddington mass accretion rates, the accretion flow is affected by advective transport of viscously generated heat \citep{1988ApJ...332..646A,1998MNRAS.297..739B} and only a fraction $\varepsilon_{\rm w}<1$ of the energy dissipated in the accretion disc is used to produce the outflow. The outflow is produced within the spherization radius $R_{\rm sp}$.
The expected (for a given $\dot{m}_0$ and $\varepsilon_{\rm w}$) mass accretion rate at the ISCO can be estimated as 
\be
\dot{M}_{\rm ISCO}=\dot{M}_0\,\frac{1-A}{1-A\,(0.4\,\dot{m}_0)^{-1/2}}, 
\ee
while the mass accretion rate at an arbitrary radial coordinate $R$ within the spherization radius can be estimated as
\be\label{eq:dotM(R)}
\dot{M}(R)=\dot{M}_{\rm ISCO}+(\dot{M}_0 -\dot{M}_{\rm ISCO})\frac{R}{R_{\rm sp}}, 
\ee
where $A\approx \varepsilon_{\rm w}(0.83-0.25\varepsilon_{\rm w})$.
The actual inner disc radius, $R_m$, depends on the mass accretion rate in the inner disc.
As a result, both the inner disc radius and the mass accretion rate there (as well as the outflow rate) are determined by a non-linear system of equations (\ref{eq:Rm}) and (\ref{eq:dotM(R)}), which can be solved numerically for given input values of $\varepsilon_{\rm w}$, $\dot{M}_0$ and $B$.

\begin{figure}
\centering 
\includegraphics[width=8.5cm]{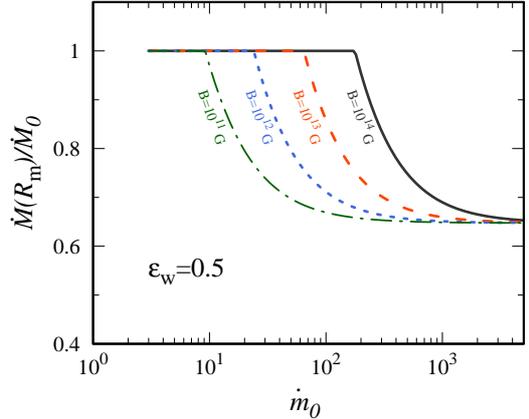} 	
\caption{The mass accretion rate reaching the inner disc radius for an accreting strongly magnetized NS, as a fraction of the accretion rate from the donor. The mass losses due to the wind from accretion disc and advection of viscously generated heat are taken into account. Different curves are given for different magnetic field strength.
The magnetic field is taken to be dominated by the dipole component. 
It is assumed that only half of the heat dissipated in the accretion disc is used to produce the outflow.
Parameters: $\Lambda=1$.
}
\label{pic:sc_Mdot2}
\end{figure}

\begin{figure}
\centering 
\includegraphics[width=8.5cm]{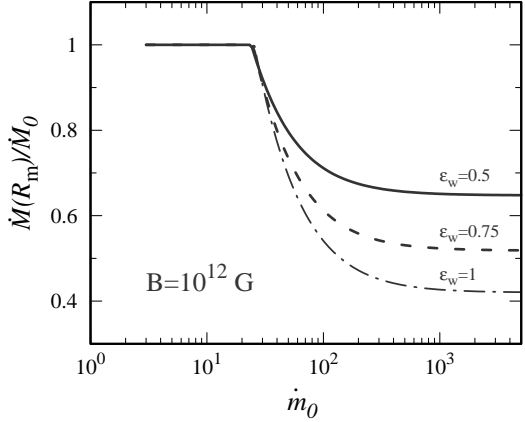} 	
\caption{The fraction of the mass accretion rate from the donor reaching the inner radius of the accretion disc, accounting for advection and mass losses due to a wind. The magnetic field at the NS surface is taken to be $10^{12}\,{\rm G}$ and assumed to be dominated by the dipole component. Different curves correspond to different fractions of viscously generated heat used to produce the outflow. Parameters: $\Lambda=1$.
}
\label{pic:sc_Mdot3}
\end{figure}

\begin{figure}
\centering 
\includegraphics[width=8.5cm]{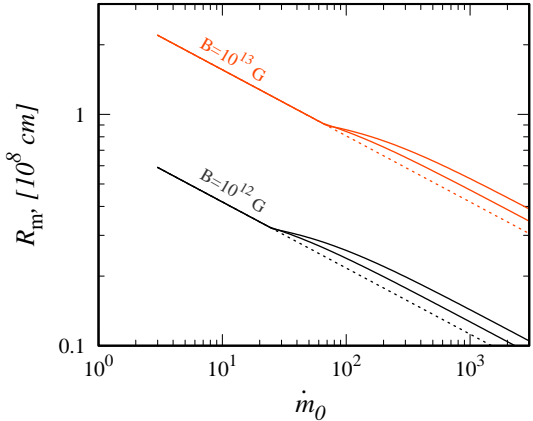} 	
\caption{The dependence of the magnetospheric radius on the mass accretion rate from the donor. 
Red and black lines are calculated for surface magnetic field strength of $10^{13}\,{\rm G}$ and $10^{12}\,{\rm G}$ respectively.
The dotted lines represent the dependencies given by equation (\ref{eq:Rm}), which do not account for mass losses from the disc.
The solid lines represent the dependencies accounting for mass losses and the advection process in the disc. 
Different solid lines correspond to different fractions of viscously generated heat used to produce the outflow: $\varepsilon_{\rm w}=0.5,\,1$ (down, up).
The magnetic field is assumed to be dominated by the dipole component.
Parameters: $\Lambda=1$.}
\label{pic:sc_Rm2}
\end{figure}

For a given mass accretion rate in the outer disc and magnetic field strength at the NS surface (assuming the field is dominated by the dipole component), we calculate the inner disc radius and mass accretion rate there. Because of the advection process, the accretion flow loses only a fraction of the mass transferred from the donor. The maximal fractional outflow rate depends on $\varepsilon_{\rm w}$ and cannot exceed $\sim 60\%$ of the initial mass inflow rate (see Fig.\,\ref{pic:sc_Mdot2},\ref{pic:sc_Mdot3}). The mass accretion rate at the inner disc radius $\dot{M}(R_{\rm m})$ and, therefore, the total mass outflow rate $(\dot{M}_0-\dot{M}(R_{\rm m}))$, strongly depend on NS magnetic field strength (see Fig.\,\ref{pic:sc_Mdot2}): in the case of extremely strong $B$-fields, an intensive outflow is possible only in the case of extreme mass accretion rate from the donor because only in this case is the spherisation radius larger than the magnetospheric radius ($R_{\rm sp}>R_{\rm m}$). 
\textit{Therefore, the detection of a strong outflow in a ULX pulsar (e.g. Kosec et al. 2018) can put an upper limit on the dipole component of NS magnetic field strength.} The correction to the inner disc radius due to mass losses from super-Eddington accretion are not dramatic and do not exceed a factor of $\sim 1.3$ (see Fig.\,\ref{pic:sc_Rm2}). Despite the possibility of significant mass losses from the disc, accretion of mass onto the NS surface dominates the total luminosity, whereas the disc itself contributes $L_{\rm disc}<{G M \dot{M}_0}/R_{\rm m}$ only.

It is worth noting that the approximations proposed by \citealt{1999AstL...25..508L} and \citealt{2007MNRAS.377.1187P}, and used here as a base for our estimations, were designed for the case of accreting BHs, i.e. they assume a zero-torque boundary condition at the ISCO 
and the energy release in accretion disc only, while in the case of accretion onto magnetized NS, the majority of the accretion luminosity is produced at the stellar surface.
In addition, the accretion discs in these papers were considered to be Keplerian, which is not necessarily the case for the inner disc regions at extreme mass accretion rates due to a possibly significant radiation pressure gradient in the accretion flow (see \citealt{1988ApJ...332..646A,2010arXiv1005.5279S}). Because the inner parts of accretion disc are expected to be radiation pressure dominated in ULX pulsars, the geometrical thickness of accretion flow is almost independent on the radial coordinate \citep{2007ARep...51..549S}. Thus, the disc is self-shielded from the central source and the mass loss rate is not affected by the energy release at the NS surface. However, the velocity of the outflow and its geometry are expected to be under influence of the energy release at the central object. 
Both the non-zero-torque inner boundary condition at the inner disc radius and deviation of the accretion flow from Keplerian velocities tend to slightly reduce the local energy release. As a result, both of these effects reduce the mass outflow rate. Therefore, although the estimates presented here should be considered approximations, we note that a more accurate treatment will increase the outflow rate for a given set of parameters, thus \textit{reducing} further the upper limit on NS dipole magnetic field strength provided by detection of an outflow.

\subsection{The optical thickness of the envelope affected by the outflow}

For the calculated inner disc radius and mass accretion rate there we can re-calculate the structure of the accretion flow covering the NS magnetosphere (i.e. the dependence of velocity and optical thickness of the material at the magnetospheric surface on the coordinate $\lambda$), modifying the calculations of \citet{2017MNRAS.467.1202M}, where the mass losses from the disc were not taken into account. 
As in \citet{2017MNRAS.467.1202M}, we make an approximation that the NS magnetic dipole is aligned with the accretion disc plane. We do not account for the centrifugal force in the envelope, which naturally arises in the reference frame co-rotating with a NS. These assumptions still allow for rough estimates for the optical thickness in the envelope (accounting for the centrifugal force results in a larger optical thickness all over the envelope, which amplifies the effects of out interest). 

\begin{figure}
\centering 
\includegraphics[width=8.5cm]{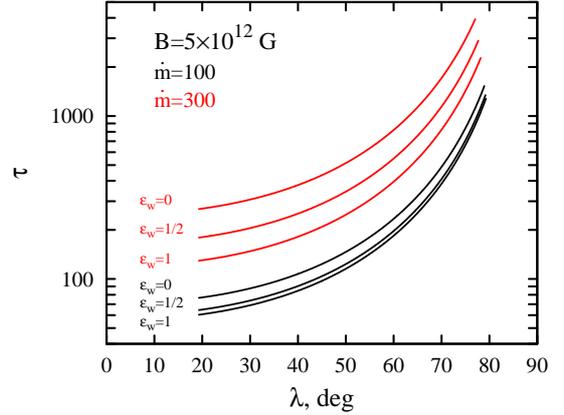} 	
\caption{The dependence of optical thickness of the envelope at the magnetospheric surface on the angular coordinate $\lambda$. $\lambda=0$ at the accretion disc plane and $\lambda= 90^\circ$ at the NS magnetic pole (see Fig. 1 in \citealt{2017MNRAS.467.1202M}).
Different lines are given for different mass accretion rates and different parameters $\varepsilon_{\rm w}$.
Parameters: $\Lambda=1$.
}
\label{pic:sc_examp_tau01}
\end{figure}

\begin{figure}
\centering 
\includegraphics[width=8.5cm]{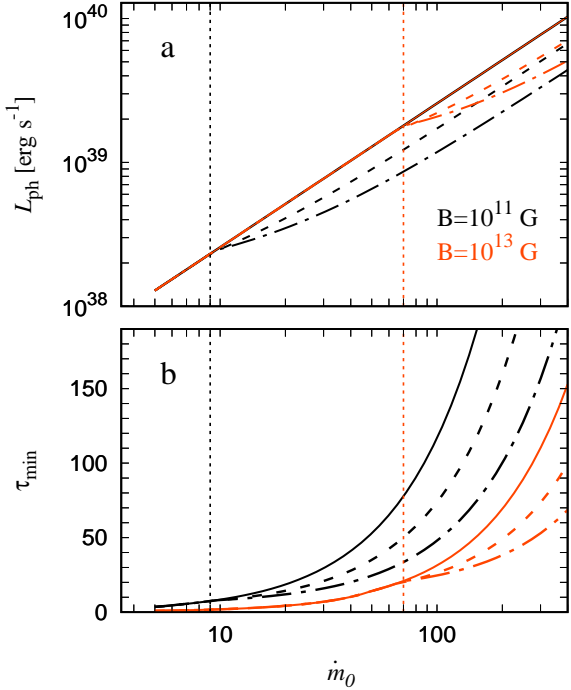} 	
\caption{Photon accretion luminosity (a) and minimal optical thickness of the envelope (b) as a function of the mass accretion rate from the donor.
Black and red lines correspond to surface $B$-field strengths of $10^{11}\,{\rm G}$ and $10^{13}\,{\rm G}$ respectively. 
Solid, dashed and dashed-dotted lines correspond to different values of wind launching efficiency: $\varepsilon_{\rm w}=0,\,0.5$ and $1$ respectively.
Both luminosity and minimal optical thickness of the envelope are affected by mass loss from the disc above a certain mass accretion rate (vertical dotted lines), which depends on the field strength.
Parameters: $\Lambda=1$. The calculations of the optical thickness did not account for centrifugal force.}
\label{pic:sc_tau_wOut}
\end{figure}

\begin{figure}
\centering 
\includegraphics[width=8.5cm]{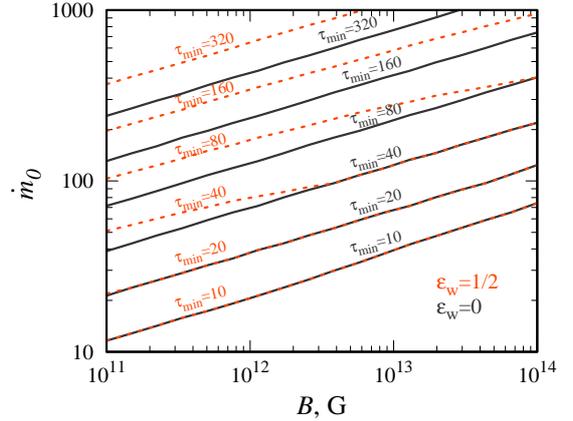} 	
\caption{Lines of constant minimal optical thickness of the envelope for different values of surface magnetic field strength and mass accretion rate from the donor. Black solid lines are calculated without accounting for mass losses from the accretion disc due to the wind, while red dashed lines are calculated accounting for these losses with the coefficient of efficiency $\varepsilon_{\rm w}=0.5$.
Parameters: $\Lambda=1$. The calculations of the optical thickness did not account for the centrifugal force.}
\label{pic:sc_tau00}
\end{figure}

The optical thickness of the accretion flow forming the envelope is determined by the mass accretion rate from the companion star (compare black and red lines in Fig.\,\ref{pic:sc_examp_tau01}), the magnetic field strength and the efficiency of the outflow launching (see Fig.\,\ref{pic:sc_examp_tau01},\,\ref{pic:sc_tau_wOut}b).
Fig.\,\ref{pic:sc_examp_tau01} represents the distribution of the optical thickness over the magnetospheric surface, where the coordinate angle $\lambda$ is measured from the equator of the magnetic dipole, i.e. $\lambda=0$ at the accretion disc plane and $\lambda= 90^\circ$ at the NS magnetic pole (see Fig. 1 in \citealt{2017MNRAS.467.1202M}).
Higher efficiency of the wind launching (i.e. larger $\varepsilon_{\rm w}$) results in smaller optical thickness all over the envelope.
The outflows reduce both the accretion luminosity of ULX pulsars (see Fig.\,\ref{pic:sc_tau_wOut}a) and the minimal optical thickness of the envelope (see Fig.\,\ref{pic:sc_tau_wOut}b) because they reduce the mass accretion rate in the disc, leading to expansion of the envelope. However, the minimal optical thickness of the envelope is still around a few tens for reasonable $B$-field strength and mass accretion rates typical for ULX pulsars, $\dot{m}\gtrsim 50$ (see Fig.\,\ref{pic:sc_tau00}).

\section{The influence of the envelope on aperiodic variability properties}
\label{sec:TransferFun}

In this section, we consider how the envelope transforms the time dependence of the flux from the NS surface as seen by the observer, using the approximation of a spherical envelope with constant optical thickness. We first consider how an infinitely short flare from the NS surface is transformed by the envelope. This transformed signal as a function of time is the impulse-response function for a signal penetrating through the envelope. The observed signal can be represented as a convolution of the initial time-dependent signal and the impulse-response function in the time domain. 
In the frequency domain, the convolution turns into multiplication of the Fourier transforms of the initial signal and the transfer function (which is the Fourier transform of the impulse-response function). Assuming a spherical envelope provides a crude approximation of the actual envelope around ULX pulsars that nonetheless provides qualitative insight into the effects arising from penetration of X-ray energy flux through the magnetosphere, which is completely covered by accreting material.
\footnote{The similar method based on the analyses of the impulse-response function and the transfer function was proposed by \citealt{1982ApJ...255..419B} to investigate reverberation process in Seyfert galaxies and quasars.}

\subsection{The time distribution of photons experiencing multiple scatterings}

Let us consider a medium with a given constant absorption coefficient $\alpha_{\rm abs}$, which determines the absorption process and is defined by the cross section of interaction $\sigma$ and the number density of particles $n$: $\alpha_{\rm abs}=\sigma n$.
In the case of our interest, the main mechanism of opacity is Compton scattering and, therefore, the cross section of interaction is given by the Thomson scattering cross section: $\sigma=\sigma_{\rm T}$, while the number density of particles is the number density of free electrons $n=n_{\rm e}$. 
As a result, the absorption coefficient $\alpha_{\rm abs}=\sigma_{\rm T}n_{\rm e}$. The mean free path length is given by $l_{\rm sc}=\alpha_{\rm abs}^{-1}$, while the mean time between scatterings is $t_{\rm sc}=l_{\rm sc}/c$, where $c$ is the speed of light. 

Let us consider an infinitely short flare in the medium at $t_0=0$.
Each photon will experience a scattering after a while. 
The specific intensity of radiation transmitted through the medium decays exponentially with the optical thickness of that medium. 
If the medium is homogeneous and the scattering cross section does not depend on the photon energy, one can introduce a typical time scale $t_{\rm sc}$ between scattering events. Then the distribution of photons over the time to the next scattering is given by
\be
I^{(1)}(t)=e^{-(t-t_0)/t_{\rm sc}}=e^{-t/t_{\rm sc}}. 
\ee
The time distribution of photons experiencing two scattering events can be calculated as:
\be\nonumber
I^{(2)}(t)=\int\limits_{0}^{t}\d t'\, I^{(1)}(t')e^{-(t-t')/t_{\rm sc}}.
\ee
More generally, the distribution of photons experiencing $n$ scattering events over the time is given by:  
\be
I^{(n)}(t)=\int\limits_{0}^{t}\d t'\, I^{(n-1)}(t')e^{-(t-t')/t_{\rm sc}}, 
\ee
which results in: 
\be\label{eq:I^n}
I^{(n)}(t)=\frac{(t/t_{\rm sc})^{n-1}}{(n-1)!}e^{-(t/t_{\rm sc})}.
\ee 
Note that the distribution is normalized to unity: $\int_0^{\infty}\d t\,I^{(n)}(t)=1$. 
Equation (\ref{eq:I^n}) describes the time distribution of a signal consisting only of photons undergoing exactly $n$ scatterings in the medium. 
Using a Fourier transform we get the equivalent distribution in the frequency domain:
\be
\overline{I}^{(n)}(\omega)=(1+\omega^2)^{-n/2}\exp\left[ -i\,n\,{\rm atan}\,\omega \right],
\ee
where $\omega$ denotes the angular frequency. 
The radiation field consists of photons which have experienced different numbers of scatterings. 
If the distribution of photons over the number of scatterings is known and given by $J_n$ with normalization 
$\sum_{n=0}^{\infty}J_n=1$, then
the photon distribution over time before detection is:
\be\label{eq:I_t}
I(t)=\sum\limits_{n=0}^{\infty}J_n I^{(n)}(t), 
\ee
where $J_n$ is the total intensity of radiation consisting of photons that have undergone exactly $n$ scatterings.
$J_n$ is defined by the geometry of the problem and does not depend on time, while $I^{(n)}(t)$ does not depend on geometry and describes the intensity as a function of time.
It is clear that the total intensity in the frequency domain is given by:
\be\label{eq:I_omega}
\overline{I}(\omega)=\sum\limits_{n=0}^{\infty}J_n \overline{I}^{(n)}(\omega). 
\ee

\subsection{Plane parallel layer}

\begin{figure}
\centering 
\includegraphics[width=8.5cm]{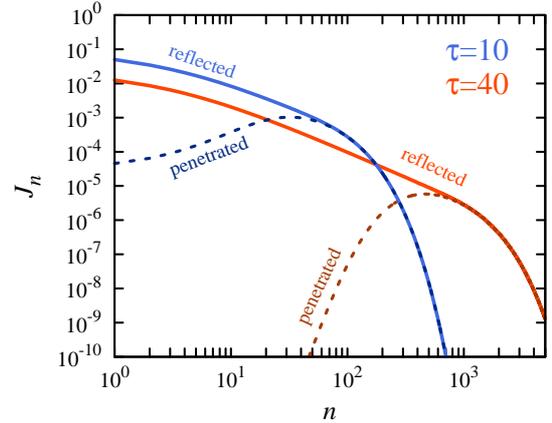}     
\caption{The distribution of photons over number of scatterings in a plane parallel layer illuminated from one side. The optical thickness of the layer was taken to be $\tau =10$ (blue) and $\tau=40$ (red). Solid and dashed lines correspond to the photons reflected back and penetrated through the layer.}
\label{pic:NumScat}
\end{figure}

Now let us consider a plane parallel layer of a given optical thickness $\tau$ and solve the radiative transfer problem numerically (see Appendix \ref{sec:App1}) to get the distributions of photons reflected from the layer and penetrating through the layer over the number of scattering events, i.e. coefficients $J_n$ in equations (\ref{eq:I_t}) and (\ref{eq:I_omega}). 
The examples of photon distributions over the number of scatterings are given in Fig.\,\ref{pic:NumScat}.
The distribution of photons reflected by the layer $J_n^{\rm in}$ (solid lines in Fig.\,\ref{pic:NumScat}) is monotonic and the majority of these photons undergo a small number of scatterings. 
The distribution of photons penetrating through the layer $J_n^{\rm out}$ has a local maximum, which corresponds to the number of scatterings necessary to penetrate through the layer (it is roughly $\sim \tau^2$, see e.g. \citealt{1979rpa..book.....R}).
The distributions are normalized as: 
\be
\sum\limits_{i=0}^{\infty}J_i^{\rm out}+\sum\limits_{i=1}^{\infty}J_i^{\rm in}=1. 
\ee
The fraction of photons that can penetrate through the layer is given by the ratio
\be 
f_{\rm out}=\left(\sum\limits_{i=0}^{\infty}J_i^{\rm out}\right)\left(\sum\limits_{i=1}^{\infty}J_i^{\rm in}\right)^{-1}.
\ee
and depends on the optical thickness $\tau$ of the layer.
In the case of $\tau\gg 1$ the fraction of penetrated photons can be approximated by (see Appendix \ref{sec:Fraction})
\be
f_{\rm out}\approx \frac{0.6}{\tau}. 
\ee
This means that the majority of photons in ULX pulsars (where the optical thickness of the envelope can be much larger than $10$, see e.g. \citealt{2017MNRAS.467.1202M}) cannot penetrate through the envelope right away and are instead reflected back into the envelope. 
The photons reflected back into the envelope cross it and then have a second chance to penetrate through and be emitted from the other side. 

\subsection{The approximation of a spherical envelope}

In order to account for multiple events, when photons cross the envelope, we consider a simple model of a geometrically thin spherical envelope of radius $R$, which is assumed to be close to the magnetospheric radius ($R\sim R_{\rm m}$), and has constant optical thickness $\tau$.
There are two time scales in the problem: $t_{\rm in}=2R/c$ - the time taken for a photon to cross the diameter of the sphere, and the typical time between scatterings in the envelope $t_{\rm sc}$, which is determined by the typical number density of electrons in the envelope.

The distribution of reflected photons over time of travel inside the spherical envelope is given by (see Appendix \ref{App:PhotonTimeDistrib}):
\be
G(t)=\frac{t c^2}{2R^2}, 
\ee
where $t\in [0, 2R/c]$ and $\int^{2R/c}_0\,\d t\,G(t)=1$. 
In the frequency domain the distribution $G(t)$ transforms into:
\be
\overline{G}(\omega)=\frac{2}{t_{\rm in}^2 \omega^2}
\left[(i t_{\rm in}\omega +1)e^{-i t_{\rm in}\omega}-1
\right]. 
\ee
The function $G(t)$ describes the time distortion of a signal due to the process of propagation inside the spherical envelope.

As soon as we know the coefficients $J_n^{\rm in}$ and $J_n^{\rm out}$, and the functions $G(t)$ and $\overline{G}(\omega)$, we can model the timing properties of a signal penetrating through the envelope. 
Let $I_{\rm ini}(t)$ be a function describing the variable brightness of a source in the center of the envelope.
Then the intensity of flux penetrating through the envelope from the \textit{first} try is: 
\be\nonumber
I_1^{\rm out}(t)= I_{\rm ini}(t) f_{\rm out}\otimes\sum\limits_{n=0}^{\infty}J_n^{\rm out}I^{(n)}(t),
\ee
while the flux {\it reflected} back into the envelope from the \textit{first} try of penetration through the envelope is:
\be\nonumber
I_1^{\rm in}(t)= I_{\rm ini}(t) (1- f_{\rm out})\otimes\sum\limits_{n=0}^{\infty}J_n^{\rm in}I^{(n)}(t),
\ee
where $\otimes$ denotes a convolution.
Then the photons reflected back into the envelope travel inside the envelope and try to penetrate through it once more. 
The intensity of flux penetrating through the envelope from the $(N+1)^{\rm th}$ try is:
\be\nonumber
I_{N+1}^{\rm out}(t)= \left[ I_N^{\rm out}(t) f_{\rm out}\otimes G(t)\right]\otimes\left[ \sum\limits_{n=0}^{\infty}J_n^{\rm out}I^{(n)}(t)\right],
\ee
The flux reflected back into the envelope from the $(N+1)^{\rm th}$ attempt to penetrate through the envelope is given by: 
\be\nonumber
I_{N+1}^{\rm in}(t)= \left[ I_N^{\rm out}(t) (1- f_{\rm out})\otimes G(t)\right]\otimes\left[ \sum\limits_{n=0}^{\infty}J_n^{\rm in}I^{(n)}(t)\right].
\ee
In the frequency domain a convolution turns to a product and the expression can be rewritten as:
\beq
\overline{I}^{\rm out}_{N+1}(\omega)&=&f_{\rm out} \overline{I}^{\rm in}_N(\omega) 
\left[ \sum\limits_{n=0}^{\infty} J^{\rm out}_n \overline{I}^{n}(\omega) \right] \overline{G}(\omega)\\
&=&\overline{I}_{\rm ini}(\omega)f_{\rm out}(1-f_{\rm out})^{N-1}\left[ \sum\limits_{n=0}^{\infty} J^{\rm out}_n \overline{I}^{n}(\omega) \right]^N \overline{G}^{N-1}(\omega)
\nonumber
\eeq
and
\beq
\overline{I}^{\rm in}_{N+1}(\omega)=(1-f_{\rm out}) \overline{I}^{\rm in}_N(\omega) 
\left[ \sum\limits_{n=0}^{\infty} J^{\rm in}_n \overline{I}^{n}(\omega) \right] \overline{G}(\omega).
\eeq
Because the flux emitted by the envelope is composed of the photons that experienced any possible number of scatterings, the flux detected by a distant observer can be represented in the frequency domain as:
\be
\overline{I}^{\rm out}(\omega)=\sum\limits_{N=1}^{\infty}\overline{I}^{\rm out}_N (\omega). 
\ee
As a result, we get an expression for the filter describing the transformation of the initial variability of X-ray energy flux.
The shape of the filter is determined by two time scales: the typical time between scattering events in the envelope $t_{\rm sc}$ and the typical time of photon crossing the envelope after multiple scattering in it $t_{\rm in}$, and 
therefore, depends on the geometrical size of the spherical envelope, the optical thickness of the envelope and its geometrical thickness. 
Because the geometrical thickness of the envelope is determined by the penetration depth of stellar magnetic field into the disc at its inner radius, which is highly uncertain (see \citealt{2014EPJWC..6401001L} for review), the time scale $t_{\rm sc}$ is not well defined. 
Thus, we consider the optical thickness $\tau$ and the ratio $t_{\rm in}/t_{\rm sc}$ as a free parameters of our model.

\begin{figure}
\centering 
\includegraphics[width=8.5cm]{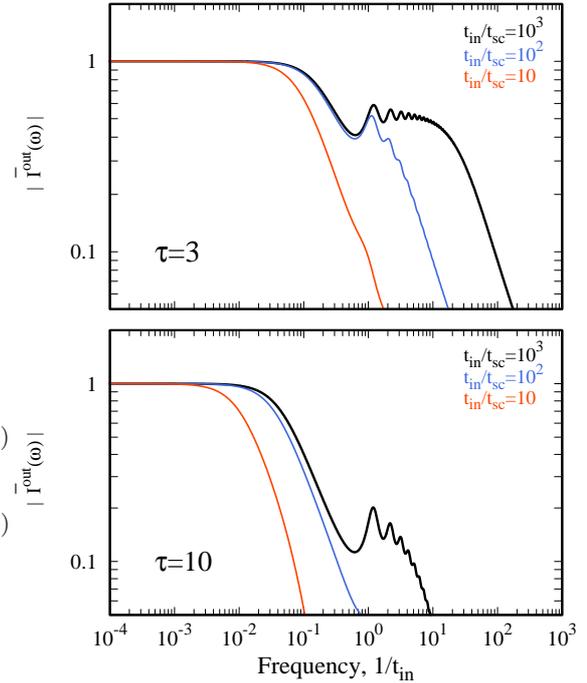} 	
\caption{The absolute value of the transfer functions $\overline{I}^{\rm out}(\omega)$ calculated for the case of multiple scatterings in the spherical envelopes of optical thickness $\tau=3$ (upper panel) and $\tau=10$ (lower panel). Different curves are given for different ratios of typical time scales $t_{\rm in}/t_{\rm sc}$: $1000$ (black), $100$ (blue) and $10$ (red).}
\label{pic:TrFunT_3_10}
\end{figure}

\begin{figure}
\centering 
\includegraphics[width=8.7cm]{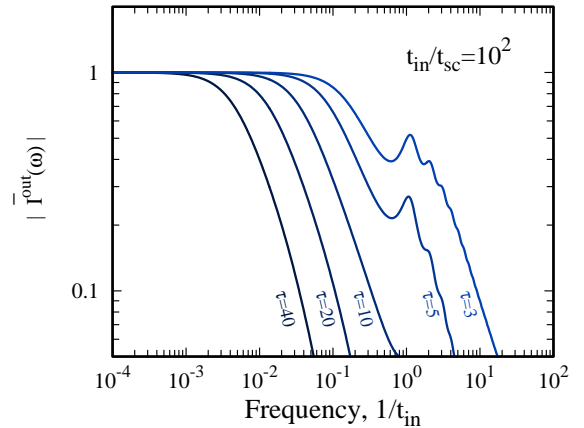} 	
\caption{The absolute value of the transfer functions $\overline{I}^{\rm out}(\omega)$ calculated for the case of multiple scatterings in the spherical envelope of different optical thickness. 
The ratio of typical time scales was fixed at $t_{\rm in}/t_{\rm sc}=100$.
One can see that the larger the optical thickness of the envelope, the stronger the suppression of variability at high frequencies.
}
\label{pic:TrFun}
\end{figure}

Examples of the filtering function are given in Fig.\,\ref{pic:TrFunT_3_10},\ref{pic:TrFun}.
One can see that the process of photon penetration through the envelope results in suppression of variability at high Fourier frequencies. 
The strength of suppression is determined by the optical thickness of the envelope: the thicker the envelope, the larger the typical number of scatterings until escape and the stronger the suppression of variability at high Fourier frequencies (see Fig.\,\ref{pic:TrFun}). 
The exact shape of the filtering function is affected by the ratio of two time scales of the problem $t_{\rm in}/t_{\rm sc}$: in the case of $t_{\rm in}\gg t_{\rm sc}$ the shape of the filtering function can be quite complicated (see Fig.\,\ref{pic:TrFunT_3_10}) because of interference between the photon energy fluxes leaving the system after different numbers of reflections inside the envelope. 
The first interference peak is located at the frequency corresponding to the typical light crossing time of the envelope. 
The thicker the envelope, the weaker the interference peaks (compare Fig.\,\ref{pic:TrFunT_3_10}a and \ref{pic:TrFunT_3_10}b).
We note that these interference peaks may be smoothed out in a calculation accounting for a non-spherical envelope with an optical depth that is not constant, but the supression of high frequency variability should be fairly robust to more sophisticated assumptions than those employed here.

Because the Comptonization process in the envelope tends to make the energy spectrum softer, one would expect time lags between hard and soft X-rays in ULX pulsars: softer X-rays  undergo a larger number of scatterings and are therefore expected to lag hard X-rays.

The approximation of a spherical envelope provides only qualitative predictions on the modifications of PDS in ULX pulsars. 
The actual shape of the envelope can be complicated and the optical thickness of the envelope is likely varies over its surface (see Fig.\,\ref{pic:sc_examp_tau01}).
Accounting for these features will result in even stronger suppression of aperiodic variability and smoothed features in PDS.
The proposed model implies that the accretion envelope is closed (i.e. there are no holes in it), otherwise the X-ray flux will escape through the holes without significant modification of its timing properties.

\subsection{\textit{STROBE-X} simulation}

We consider the prospects for observing the power spectral signatures of the envelope predicted in Figs \ref{pic:TrFunT_3_10} and \ref{pic:TrFun}. Due to large source distances and consequently relatively low observed flux, aperiodic variability analysis of ULX pulsars (and ULXs in general) is observationally very challenging (although there are papers which have studied the noise processes in ULXs; see e.g. \citealt{2009MNRAS.397.1061H,2015MNRAS.447.3243M}). 
Constraints on the intrinsic high frequency variability is particularly challenging due to the effects of Poisson noise. X-ray observatories with higher effective area than those currently in operation are therefore required. To illustrate this point, let us consider the signal to noise ratio of a measured power spectrum
\begin{equation}
[S/N]_p = \frac{ P~\sqrt{T \Delta} } { P + 2(s+b)/s^2},
\label{eqn:SN}
\end{equation}
where $s$ and $b$ are respectively the source and background count rates, $T$ is the exposure time and $P$ is the intrinsic power (in units of squared fractional rms per Hz) averaged over a frequency range of width $\Delta$ (see e.g. \citealt{1989ARA&A..27..517V}). The source count rate is thus very important to suppress Poisson noise. The expected source count rate of a typical ULX pulsar (NGC 7793 P13; calculated from the spectral model of \citealt{2018ApJ...857L...3W}) is $\sim 0.85~{\rm c~s}^{-1}$ for the \textit{XMM-Newton} EPIC-pn and $\sim 1.3~{\rm c~s}^{-1}$ for \textit{NICER}, whereas the large area detector (LAD) and X-ray concentrator array (XRCA) instruments proposed to be on board \textit{STROBE-X} would measure $\sim 39~{\rm c~s}^{-1}$ and $\sim 13~{\rm c~s}^{-1}$ respectively. Although the LAD will measure a much higher source count rate, the XRCA will actually achieve the best signal to noise ratio ($[S/N]_p\approx 23.5$ for XRCA and $[S/N]_p\approx 2.2$ for the LAD, assuming $P=0.01 {\rm rms}^2/{\rm Hz}$ and $\Delta=1$ Hz) because it's \textit{background} count rate is so much lower ($\sim 2.2~{\rm c~s}^{-1}$ for the XRCA $\sim 1,486~{\rm c~s}^{-1}$ for the LAD; Ray et al 2018). The signal to noise for the cross-spectrum between the XRCA and LAD light curves is (e.g. \citealt{1994ApJ...421..738V})
\begin{equation}
[S/N]_c = \frac{ 2 P~\sqrt{T \Delta} } { \sqrt{ (P + 2(s_l+b_l)/s_l^2) (P + 2(s_x+b_x)/s_x^2) } },
\label{eqn:SN}
\end{equation}
where subscript $l$ ($x$) corresponds to the LAD (XRCA). For the same parameters as previously considered, we find $[S/N]_c\approx 14.5$. We therefore conclude that the best diagnostic in this case is the XRCA power spectrum.

\begin{figure}
\centering 
\includegraphics[width=8.5cm]{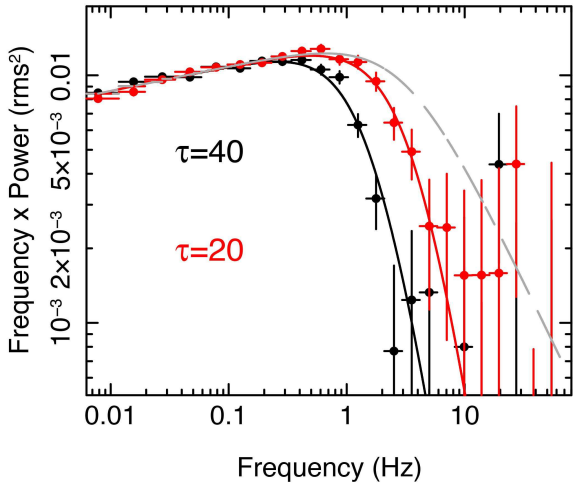} 
\vspace{-0.3cm}
\caption{Power spectrum for a simulated $200$ ks observation of a typical ULX pulsar with the XRCA from the proposed NASA mission \textit{STROBE-X}. For the intrinsic power spectrum (grey dashed line), we use a model for sub-critical X-ray pulsars (discussed in the text). For the envelope, we use the same parameters as in Fig \ref{pic:TrFun}, with the optical depth value as labelled and assuming $1/t_{\rm in}=300$ Hz (corresponding to $R_m \sim 10^{8}$ cm and $B\sim 10^{12}$ G). We see that the influence on the power spectrum of the optically thick envelope could be constrained with such an observation.}
\label{pic:strobex}
\end{figure}

Fig.~\ref{pic:strobex} shows the power spectum of a $200$ ks simulated XRCA observation. We assume that the intrinsic power spectrum is given by a bending power-law, $P_{\rm ini}(\nu) = x^{-\lambda} (1+x^\kappa)^{(\lambda-\zeta)/\kappa}$, where $x=\nu/\nu_{\rm br}$ and $\omega = 2\pi \nu$. This assumes a model whereby variability is produced throughout the accretion disc predominantly at the dynamo timescale, such that the break frequency coincides with the dynamo frequency at the magnetospheric radius (Mushtukov et al in prep). We set $\nu_{\rm br}=3$ Hz, $\zeta=2$, $\kappa  = 1.5$ and $\lambda=0.9$, which gives a power spectrum consistent with those observed from Galactic X-ray pulsars (e.g. \citealt{2009A&A...507.1211R}). Here, power is in units of squared fractional variability amplitude per Hz. We calculate the transfer function of the envelope using the same parameters as for Fig.\, \ref{pic:TrFun}, with the optical depth value as labelled and assuming $1/t_{\rm in}=300$ Hz (corresponding to $R_{\rm m} \sim 10^{8}$ cm and $B\sim 10^{12}$ G). We see that change in power spectral slope at high frequencies introduced by scattering in the envelope can be constrained for the two optical depths considered. In particular, we note that in the absence of an optically thick envelope the break frequency, thought to correspond to the dynamo timescale at the magnetospheric radius, is expected to and observed to increase with source flux ($\nu_{\rm br}\propto L^{3/7}$), since $R_{\rm m}$ will decrease with increasing accretion rate \citep{2009A&A...507.1211R}. However, the power spectral break introduced by the envelope should move to \textit{lower} frequency as the source flux increases and the optical depth of the envelope consequently increases. Therefore, our model makes the prediction that the power spectral break frequency should decrease with luminosity for ULX pulsars, in contrast to what is observed for normal X-ray pulsars. The interference feature at higher frequencies cannot be constrained even by \textit{STROBE-X}.

\section{Summary and Discussion}
\label{discussion}

We have constructed a simple model of mass transfer in ULXs powered by accretion onto magnetized NSs and investigated the influence of the optically thick envelope on timing properties of ULX pulsars.

We show that ULX pulsars may lose a significant fraction of accreting material due to the winds launched from the surface of the supper-Eddington accretion disc.
Because of the advection process, ULXs cannot lose more than about 60 percent of the initial mass accretion rate in the outflow (see Fig.\,\ref{pic:sc_Mdot3}). 
The rest of the material forms an envelope covering the magnetosphere of the NS and provides the major fraction of the accretion luminosity.

The outflow rate is determined by the mass inflow rate from the donor star and strength of the dipole component of the NS magnetic field.
A strong magnetic field disrupts the accretion disc flow at the magnetospheric radius and restricts mass losses below a certain value. 
In particular, in the case of an extremely strong dipole component (corresponding to $B>few\times 10^{13}\,{\rm G}$ at the NS surface), the outflow rate is expected to be negligibly small because the disc is truncated at large distances from the central object where the accretion flow is still sub-critical (i.e. $R_{\rm m}>R_{\rm sp}$, see Fig.\,\ref{pic:sc_Mdot2}).
Thus, we argue that a detection of a strong
outflow from a ULX pulsar (e.g. Kosec et al. 2018) can be considered as evidence of a relatively weak dipole component of the NS magnetic field.

It is remarkable that an outflow has been recently discovered in the ULX pulsar in NGC~300 \citep{2018MNRAS.479.3978K}, with an X-ray luminosity of $L\approx 4.7\times 10^{39}\,\ergs$ and surface magnetic field estimated as $B\sim 3\times 10^{12}\,{\rm G}$ on the base of the \citealt{1979ApJ...234..296G} torque model and the detected spin period derivative \citep{2018MNRAS.476L..45C}. 
\footnote{This magnetic field strength is also consistent with the ``implied" presence of a spectral feature that might be a cyclotron scattering line \citep{2018ApJ...857L...3W}.}

The restrictions on the dipole component of the magnetic field, however, do not exclude the possibility of strong non-dipole components.
Relatively weak dipole and strong non-dipole components of the magnetic field in ULX pulsars were already proposed to explain observational data in a few ULX pulsars \citep{2017Sci...355..817I,2017A&A...605A..39T,2018MNRAS.479L.134T}.

The undetected iron lines in the energy spectra of ULXs \citep{2013ApJ...773L...9W} can be considered as indirect evidence of strong outflows in ULX pulsars, where the accretion disc is shielded from the central source by the outflow. 
Additionally, in the case of a conical geometry of the outflow, the ionization state of the outflowing material is probably high enough to hinder detection of iron lines \citep{2015MNRAS.447.3243M}.

Outflows from the disc in ULX pulsars may affect the visibility of these sources, making them detectable from certain directions only
\citep{2007MNRAS.377.1187P,2009MNRAS.393L..41K,2015MNRAS.447.3243M}.
However, because the outflows should be strongly influenced by radiative stress from the central object/envelope, the opening angle of the cone where the central source is visible for a distant observer is expected to be above $\sim 60^\circ$, which is much larger than the opening angles expected in ULXs powered by BHs, but comparable to the angles obtained in numerical simulations of super-Eddington accretion onto NSs with low magnetic fields \citep{2018ApJ...853...45T}.

Despite the possibility of strong mass losses, the envelope forming at the magnetosphere of the NS tends to be optically thick in the case of mass accretion rates typical for ULXs ($\dot{M}\sim 100\dot{M}_{\rm Edd}$, see Fig.\,\ref{pic:sc_tau00}). 
The envelope reprocesses X-ray photons and affects spectral, polarization and timing properties of ULX pulsars.
In particular, the envelope modifies the properties of the broadband aperiodic variability because multiple scatterings of X-ray photons in the envelope result in a large photon escape time and therefore strong suppression of the aperiodic variability in X-rays at high Fourier frequencies.
In that sense, the envelope plays the role of a low-pass filter. 
The strength of suppression is determined by the optical depth of the envelope (see Fig.\,\ref{pic:TrFun}) and, therefore, by the mass accretion rate reduced by outflows from larger radii.
The modification of the initial power density spectrum by multiple scatterings in the envelope is described by the transfer function, which we have calculated numerically for a simplified geometry of the envelope (see Section \ref{sec:TransferFun} and Fig.\,\ref{pic:TrFunT_3_10},\,\ref{pic:TrFun}).
The absolute value of the transfer function tends to unity at the low frequency limit, which corresponds to unsuppressed variability. 
At high frequencies the absolute value of the transfer function decreases rapidly, which corresponds to strong suppression of the variability.


\section*{Acknowledgements}

AAM was supported by the Netherlands Organization for Scientific Research (NWO). 
AI acknowledges support from the Royal Society. 
MM appreciates support via an STFC Rutherford fellowship.
This research was also supported by COST Action PHAROS (CA16214)
and the Russian Science Foundation grant 14-12-01287.

\bibliographystyle{mnras}

\appendix

\section{Numerical solution of radiative transfer equation}
\label{sec:App1}

Let us consider plane parallel homogeneous layer with the borders at the  coordinates $x_1$ and $x_2$ and an absorption coefficient $\alpha_\nu$. 
The optical thickness of the layer is $$\tau_\nu=\alpha_\nu(x_2-x_1).$$
The equation of radiative transfer describing the specific intensity $I_\nu$ can be written as
\be 
\cos\theta\, \frac{\d I_\nu (x,\theta)}{\d x}=-\alpha_\nu(x,\theta) I_\nu(x,\theta)+\varepsilon_\nu(x,\theta)+\varepsilon_\nu^{(0)}(x,\theta),
\ee
where $I_\nu(x,\theta)$ is the intensity at a given coordinate $x$ and direction $\theta$, $\varepsilon_\nu^{(0)}(x,\theta)$ is the emission coefficient describing the initial sources of radiation, $\varepsilon_\nu(x,\theta)$ is the emission coefficient due to multiple scatterings in the layer. 
The absorption coefficient $\alpha_\nu$ is determined both by processes of true absorption and scattering.
If there is no true absorption and scattering is monochromatic and isotropic the emission coefficient is determined by absorption coefficient and local intensity of radiation:
\be
\varepsilon_\nu (x,\theta)=\frac{\alpha_\nu}{2}\int\limits_0^\pi \d\theta' \sin\theta' I_\nu(x,\theta').
\ee
Now we can calculate step by step intensities and emission coefficients. 
The intensity of radiation, which has already undergo $i$ scattering events is
\be\label{eq:intens01}
I^{(i)}_\nu(x,\theta)=\int\limits_{x_1}^{x_2}\d x'\, \frac{\varepsilon^{(i)}_\nu(x',\theta)}{\cos\theta} 
\exp\left[-\frac{\alpha_\nu |x-x'|}{\cos\theta}\right],
\ee
where $\varepsilon^{(i)}_\nu$ is the source function of photons which have undergo \textit{exactly} $i$ scattering events.
The source function is given by
\be
\varepsilon^{(i+1)}_\nu(x,\theta)=\frac{\alpha_\nu}{2}\int\limits_{0}^\pi \d\theta'\sin\theta' I^{(i)}_\nu(x,\theta').
\ee
The total local intensity is composed of intensities if photons, which undergo different number of scatterings and given by
\be
I_\nu(x,\theta)=\sum\limits_{n=0}^{\infty}I^{(n)}_\nu(x,\theta), 
\ee
while the total emission coefficient is given by
\be
\varepsilon_\nu(x,\theta)=\sum\limits_{n=0}^{\infty}\varepsilon^{(n)}_\nu(x,\theta), 
\ee
where the initial emission coefficient $\varepsilon^{(0)}_\nu(x ,\theta)$ is given and the other emission coefficients have to be calculated.

\section{The distribution of photons over the time of travel within the spherical envelope}
\label{App:PhotonTimeDistrib}

Let us consider a photon reflected back into the spherical envelope.
The length of the photon trajectory inside the envelope is
\be 
l=2R\cos\theta, 
\ee
where $R$ is the radius of the envelope and $\theta\in[0;\pi/2]$ is the angle between the photon momentum after reflection and the local normal to the envelope at the point of reflection. 
The travel time of a photon inside the envelope is given by $t=l/c$.
In the case of isotropic reflection, when the intensity of the reflected radiation does not depend $\theta$, the photon distribution over the angle $\theta$ is given by
\be
f_{\theta}=\frac{\d N}{\d\theta}=2\sin\theta\cos\theta.
\ee
Thus, the distribution of photons over the travel time within the envelope is given by
\be 
f_{t}=\frac{\d N}{\d t}=f_{\theta}\frac{\d\theta}{\d t}=\frac{t\,c^2}{2R^2}.
\ee

\section{The fraction of photons penetrating through the layer: the analytical estimation}
\label{sec:Fraction}

Let us consider a plane-parallel layer of optical thickness $\tau_0$ illuminated from aside and make an estimation of the fraction of radiation penetrating through.
The photons are interacting with the matter inside the layer due to the scattering, which is assumed to be isotropic. 
The angle between the momentum of incident radiation and normal to the layer is taken to be $\arccos\zeta$, while the flux crossing the area oriented perpendicular to the photon momentum is $\pi S$. 
Thus, the photon energy flux reaching the layer is $E_{\rm inc}=2\pi S\zeta$. 
The flux reflected from and penetrated through the layer can be repesented by \citep{1969rtsc.book.....I}
\beq\label{eq:coeffApp1}
E_{\rm ref}=2\pi S\zeta\intl_0^1\rho(\eta,\zeta,\tau_0)\eta\,\d\eta,
\eeq
\beq
\label{eq:coeffApp2}
E_{\rm pen}=2\pi S\zeta\intl_0^1\sigma(\eta,\zeta,\tau_0)\eta\,\d\eta,
\eeq
respectively,
where $\rho(\eta,\zeta,\tau_0)$ and $\sigma(\eta,\zeta,\tau_0)$ are coefficients of reflection and penetration.

In the case of large optical thickness of the layer $\tau_0\gg 1$ we can use the asymptotic expressions for the coefficients of reflection and penetration (\ref{eq:coeffApp1},\ref{eq:coeffApp2}):
\beq\label{eq:coeffApp3}
\rho(\eta,\zeta,\tau_0)&=&\frac {\lambda}{4}\frac {\varphi(\eta)}{1-k\eta}
\frac {\varphi(\zeta)}{1-k\eta}\nonumber \\
&&\left[\frac {1+k^2\eta\zeta}{\eta+\zeta}-
\frac {k}{\tanh k(\tau_0+2\tau_{\rm e})}\right],\\
\sigma(\eta,\zeta,\tau_0)&=&
\frac {\lambda}{4}\frac {\varphi(\eta)}{1-k\eta}\frac {\varphi(\zeta)}{1-k\eta}
\frac {k}{\sinh k(\tau_0+2\tau_{\rm e})},
\eeq
where $\lambda\in[0;1]$ is the probability of photons to survive in a single scattering event,
$\varphi(\eta)$ is the Ambartsumian's function, $\tau_{\rm e}$ is the extrapolated length:
\be
\tau_{\rm e}=\frac {1}{2k}\ln\left(2\varphi^2(1/k)\frac {\lambda-1+k^2}{1-k^2}\right),
\ee
and $k$ is a solution of characteristic equation 
\be
\frac {\lambda}{2k}\ln\frac {1+k}{1-k}=1.
\ee

In the particular case of pure scattering $\lambda=1$, $k=0$, $\tau_{\rm e}\approx 0.71$ and the expressions (\ref{eq:coeffApp3}) can be simplified:
\beq
\rho(\eta,\zeta,\tau_0)&=&\frac {\varphi(\eta)\varphi(\zeta)}{4}\left(
\frac{1}{\eta+\zeta}-\frac{1}{\tau_0+2\tau_{\rm e}}\right),\\
\sigma(\eta,\zeta,\tau_0)&=&\frac {\varphi(\eta)\varphi(\zeta)}{4}
\frac {1}{\tau_0+2\tau_{\rm e}}.
\eeq
In order to get the total photon energy flux reflected from and penetrated through the layer we have to average over the angular distribution of incident radiation and calculate the integrals:
\be
\frac {\lambda}{2}\intl_0^1\frac {\varphi(\eta)}{1-k\eta}\d\eta=1,\quad
\frac {\lambda}{2}\intl_0^1\frac {\varphi(\eta)}{1-k\eta}\eta\d\eta=
\frac {\sqrt {1-\lambda}}{k}.
\ee
In the case of $\lambda\to 1$ (which corresponds to the pure scattering) the last integral turns to $\sqrt{3}$.
Therefore, in the simplest case of pure scattering and isotropic distribution of initial radiation we get the following approximate expression for the fraction of radiation penetrating through the layer of optical thickness $\tau_0$:
\be
f_{\rm out}=\intl_0^1\eta\d\eta\intl_0^1\d\zeta\sigma(\eta,\zeta,\tau_0)=
\frac{1}{\sqrt{3}}\frac{1}{\tau_0+1.42}.
\ee

\bsp	
\label{lastpage}
\end{document}